\documentclass[12pt,preprint,number,sort&compress]{elsarticle}
\usepackage{hyperref}
\usepackage[latin9]{inputenc}
\usepackage{amsmath}
\usepackage{amssymb}
\usepackage{graphicx}
\usepackage{esint}

\makeatletter

\providecommand{\tabularnewline}{\\}

\usepackage{psfrag}
\usepackage{url}
\usepackage{lineno}
\usepackage{array}
\usepackage{epsfig}
\usepackage{multicol}
\usepackage{times}
\usepackage{picins}

\journal{Physical Communication}

\begin{document}
\begin{frontmatter}

\title{Multi-channel Sensing And Resource Allocation in Energy Constrained
Cognitive Radio Networks}

\author{Kedar Kulkarni}

\ead{kulkarni@iitk.ac.in}

\author{Adrish Banerjee\corref{cor1}}

\ead{adrish@iitk.ac.in}

\address{Department of Electrical Engineering, Indian Institute of Technology,
Kanpur, 208016, India}

\cortext[cor1]{Corresponding author}
\begin{abstract}
We consider a cognitive radio network in a multi-channel licensed
environment. Secondary user transmits in a channel if the channel
is sensed to be vacant. This results in a tradeoff between sensing
time and transmission time. When secondary users are energy constrained,
energy available for transmission is less if more energy is used in
sensing. This gives rise to an energy tradeoff. For multiple primary
channels, secondary users must decide appropriate sensing time and
transmission power in each channel to maximize average aggregate-bit
throughput in each frame duration while ensuring quality-of-service
of primary users. Considering time and energy as limited resources,
we formulate this problem as a resource allocation problem. Initially
a single secondary user scenario is considered and solution is obtained
using decomposition and alternating optimization techniques. Later
we extend the analysis for the case of multiple secondary users. Simulation
results are presented to study effect of channel occupancy, fading
and energy availability on performance of proposed method.\end{abstract}
\end{frontmatter} 

\section{Introduction}

Cognitive radio (CR) facilitates efficient spectrum use of current
licensed spectrum that is highly underutilized and is considered as
a potential solution to the problem of spectrum scarcity \cite{FCC,Haykin}.
In CR networks, secondary users (SU) opportunistically access spectrum
allocated to licensed or primary users (PU) in such a way that quality
of service (QoS) requirements of PUs are satisfied. For this purpose,
SUs periodically sense the spectrum for presence of PUs. While many
spectrum sensing techniques exist, energy detection method is widely
used due to its low complexity and easy implementation \cite{Yucek,Sens_survey_Sun,Digham_energy_det}
and is optimal when form of signal to be detected is unknown \cite{Urkowitz_energy_det}.
SU transmits in a channel only if the channel is sensed to be vacant.
This method of spectrum access is widely known as interweave-mode
\cite{Goldsmith2}. Due to channel fading and noise, spectrum sensing
may result in missed detection or false alarm. Longer sensing periods
lead to better sensing performance, but at the cost of reduced transmission
time as a node cannot transmit and sense simultaneously. This sensing-throughput
tradeoff necessitates selection of optimal sensing time to maximize
SU throughput while sufficiently protecting PU \cite{Liang_zheng}.

Sensing throughput tradeoff where SU determines optimal sensing time
has been studied under various PU QoS constraints such as fixed target
detection probability \cite{Liang_zheng,sens_tradeoff_Peh}, collision
probability constraint \cite{sens_tradeoff_zarrin} and PU outage
constraint \cite{Sens_thru_Juarez}. Kaushik et al. \cite{Sens_tradeoff_Kaushik} studied effect
of estimation time on the tradeoff considering PU signal of unknown
power. When multiple SUs are present,
better sensing performance can be achieved in less time using cooperative
sensing. In \cite{sens_tradeoff_Peh} and \cite{Sens_thru_Juarez}, authors optimized
sensing time in cooperative sensing framework assuming availability
of a single PU band. Pei et al. \cite{sens_thru_Pei} considered
multiple PUs multiplexed using orthogonal frequency division multiple
access (OFDMA) and a SU equipped with wideband antenna, which enabled
simultaneous sensing of all PU channels. The authors determined optimal
sensing time to achieve given target detection probability and proposed
power allocation method to maximize SU throughput. Using the same
model, Sharkasi et al. \cite{sens_tradeoff_sharkasi} studied sensing
throughput tradeoff under PU outage constraint. In practice, maximum
bandwidth that can be scanned by SU is limited by its radio-frequency
(RF) frontend and analog-to-digital converter (ADC) sampling unit.
For a SU device having narrowband antenna or low sampling rate, simultaneous
sensing of all bands in a wideband spectrum is not possible, prompting
SU to optimally select sensing and transmission time in each PU band.
In this work, we aim to address this multi-channel sensing-throughput
tradeoff.

In energy harvesting (EH) wireless networks, users are often energy
constrained \cite{Paradiso,EH_Lu}. In this case, in addition to
tradeoff arising from sensing and transmission time, tradeoff in energy
becomes critical. In energy constrained CR networks, as sensing time
increases, more energy is used in sensing, leaving less energy available
for transmission. Considering EH-CR network, Park et al. \cite{EH_Park}  determined
optimal sensing threshold for SU under energy causality and collision
probability constraints. In \cite{sens_eff_Wang},
authors found optimal sensing time to minimize average energy cost
under constraints on SU transmission rate. Yin et al. \cite{sens_E_tradeoff_Yin}
divided SU frame duration in three parts for--- harvesting, sensing
and transmission--- and proposed optimal time division to maximize
SU throughput under a fixed target detection probability. A fractional
programming framework was proposed in \cite{sens_eff_Wu} to find
optimal sensing time and power allocation to maximize energy efficiency
of SU. In \cite{sens_eff_Hoang,Sens_Eff_Sultan,sens_eff_Park}, authors
proposed energy efficient dynamic control policies using Markov decision
process (MDP) approach where SU can choose to stay silent, carry out
sensing or transmit based on its belief about PU occupancy. MDP based
techniques have high computational complexity and require knowledge
of transition probabilities between different PU occupancy states.
In practice, information of state transition probabilities is not
readily available due to sparse spectrum activity over long term.
Existing spectrum availability studies only document duty cycle of
a channel which is the probability of a channel being occupied by
a PU \cite{Spectrum_survey_Chakraborty,Spectrum_survey_Das,Spectrum_survey_Lehtomaki,Spectrum_survey_Xue}.
Also, the works mentioned above considered CR systems with a single
PU channel with fixed target detection probability as PU's QoS criteria.
Availability of multiple PU channels poses a challenge as SU has to
allocate available time and energy appropriately in sensing and transmission
tasks in each channel.

In multi-channel environment, if channel conditions are such that
the channel does not yield good throughput, SUs should not transmit
in the channel. Hence SUs should not be required to sense it. Further,
choice of channels for sensing and transmission can be made based
on occupancy probability of the channel. To maximize average SU throughput,
SU must appropriately allocate limited time and energy for tasks of
sensing and transmission in each channel. In this paper, we address
the problem of finding optimal sensing time and power allocation in
a multi-channel PU environment where channels have to be sensed sequentially
such that expected bit-throughput of SU is maximized in a given duration.
We consider two main constraints--- average rate constraint of PU
to maintain QoS of PU and total energy constraint that results from
limited energy availability. Our contribution in this work is as follows. 
\begin{itemize}
\item We first consider a single SU case and formulate the joint sensing
time-energy-throughput tradeoff problem to maximize aggregate average
bit-throughput of SU. The optimization problem is a non-convex one.
We decompose the problem in subproblems with separable objectives.
We propose sensing and resource allocation (SRA) method which iteratively
solves the subproblems and finds optimal sensing time, transmission
time and transmission energy for each channel. 
\item We then extend SRA method for multiple SU scenario to maximize sum-throughput
of SU network. 
\item We present numerical results to study performance of proposed approach
under various channel and energy availability conditions. We also
compare SRA with heuristics based best channel selection (BCS) and
proportional energy-time allocation (PETA) methods. 
\end{itemize}
Rest of the paper is organized as follows. In Section \ref{sec:SysMod_Chap4},
system model is presented and optimization problem is formulated.
In Section \ref{sec:SRA_chap4}, we propose SRA and find solution
to the optimization problem. In Section \ref{sec:SRA_multi_chap4},
we propose SRA for the multiple SU case. Simulation results are presented
in Section \ref{sec:Simulation_chap4}. We conclude in Section \ref{sec:Conclusion_chap4}.

\section{System model and problem formulation\label{sec:SysMod_Chap4}}

\begin{figure}
\centering

\includegraphics[scale=0.19]{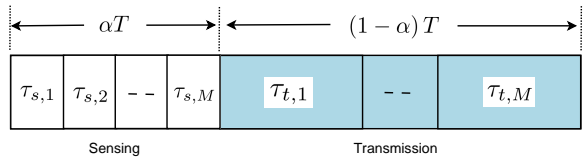}

\protect\protect\protect\caption{Frame structure for multi-channel sensing and transmission\label{fig:frame structure}}
\end{figure}

Initially we consider a cognitive radio system with one SU that opportunistically
accesses PU spectrum of $M$ non-overlapping narrowband channels of
equal bandwidth. The model for multiple SU scenario is explained later
in Section \ref{sec:SRA_multi_chap4}. PU and SU follow time slotted
synchronous communication with frame duration $T$ \cite{Zhao}.
PU is active in $i$th channel with occupancy probability $\pi_{1,i},\,i=1,\,2,\dots,\,M$.
Thus, probability of $i$th channel being vacant is $\pi_{0,i}=1-\pi_{1,i}$.
SU has a-priori knowledge of channel occupancy probabilities which
can be obtained by observing the spectrum for long duration or from
existing spectrum database \cite{Spectrum_survey_Chakraborty,Spectrum_survey_Xue,Spectrum_karandikar}.
All channels between different source-destination pairs are independent
Rayleigh block fading, that is, channel gains remain constant in one
frame and vary independently from frame to frame. Instantaneous channel
gain of SU source to SU destination link on $i$th channel is denoted
as $g_{i},\,i=1,\,2,\dots,\,M$. Noise at SU receiver is additive
white Gaussian (AWGN) with variance $\sigma_{\mathcal{N}}^{2}$. PU
transmit power in each channel is $p_{PU}$. In Table \ref{tab:Key-notations-used},
we list key notation used in this paper.

\begin{table}
\centering \protect

\begin{tabular}{|c|l|}
\hline 
$\pi_{0,i}$  & Probability of $i$th PU channel being vacant\tabularnewline
$g_{i}$  & Channel power gain of SU-SU link on $i$th channel\tabularnewline
$\sigma_{\mathcal{N}}^{2}$  & AWGN noise power\tabularnewline
$\tau_{s,i}$  & Sensing time allocated to $i$th channel\tabularnewline
$\tau_{t,i}$  & Transmission time allocated to $i$th channel\tabularnewline
$p_{t,i}$  & SU transmit power in $i$th channel\tabularnewline
$p_{s}$  & Sensing power\tabularnewline
$f_{s}$  & Sampling frequency\tabularnewline
$T$  & Frame time\tabularnewline
\hline 
\end{tabular}

\protect\protect\caption{Key notation used in this paper\label{tab:Key-notations-used}}
\end{table}

\subsection{Sensing and spectrum access}

SU is equipped with a single narrowband antenna that limits the sensing
capability to one channel at a time. At the beginning of each frame,
SU performs spectrum sensing using energy detection with sampling
frequency $f_{s}$. Sensing takes place at a constant power $p_{s}$
\cite{sens_eff_Wang}. SU senses $i$th PU channel for time $\tau_{s,i}$.
Assuming PU signal to be complex valued phase-shift keying (PSK) signal, we write
detection probability $P_{d,i}$ and false alarm probability $P_{f,i}$ as \cite{Liang_zheng}

\begin{equation}
P_{d,i}=\mathcal{Q}\left(\left(\frac{\epsilon_{i}}{\sigma_{\mathcal{N}}^{2}}-\gamma_{i}-1\right)\sqrt{\frac{f_{s}\tau_{s,i}}{2\gamma_{i}+1}}\right),\label{eq:Pd_basic}
\end{equation}
\begin{equation}
P_{f,i}=\mathcal{Q}\left(\left(\frac{\epsilon_{i}}{\sigma_{\mathcal{N}}^{2}}-1\right)\sqrt{f_{s}\tau_{s,i}}\right),\label{eq:Pf_basic}
\end{equation}
where $\epsilon_{i}$ is the detection threshold and $\gamma_{i}$
is the average PU signal-to-noise ratio (SNR) received at SU source
over $i$th channel. For a target detection probability $P_{d,i}$,
we can write false alarm probability as \cite{Liang_zheng} 
\begin{equation}
P_{f,i}=\mathcal{Q}\left(\sqrt{2\gamma_{i}+1}\mathcal{Q}^{-1}\left(P_{d,i}\right)+\gamma_{i}\sqrt{f_{s}\tau_{s,i}}\right).\label{eq:Pf_mod}
\end{equation}
Note that depending on channel occupancy probabilities, channel conditions
and available energy, SU may not sense a PU channel, which results
in $\tau_{s,i}=0$. After the sensing phase is over, for the remaining
frame duration, SU transmits in vacant PU channels (interweave mode)
with appropriate power so as maximize bit-throughput. For $i$th channel,
time spent in transmission and transmit power are denoted as $\tau_{t,i}$
and $p_{t,i}$ respectively. The frame structure of SU is shown in
Fig. \ref{fig:frame structure}.

\subsection{System constraints\label{sub:System-constraints}}

\subsubsection{Primary rate constraint}

Quality of service (QoS) criterion of $i$th PU demands that the PU
should be able to transmit $\bar{\mathcal{B}}_{p,i}$ bits on average
in each frame duration. If SU correctly senses a channel as active,
there is no interference with the PU transmission. In this case, average
number of bits transmitted by $i$th PU is given by $T\mathbb{E}\left[\mathcal{R}_{p,i}\right]$
where $\mathbb{E}\left[\mathcal{R}_{p,i}\right]$ is the average transmission
rate that depends on PU source-PU destination link. In case of missed
detection, SU transmits and interferes with $i$th PU for time $\tau_{t,i}$.
There is no interference to the PU transmission for time $\left(T-\tau_{t,i}\right)$.
We consider a strong interference channel between SU and PU. Thus,
transmission rate achieved under interference is negligible. Then
average number of bits transmitted in a frame by $i$th PU is
\begin{equation}
\mathcal{B}_{p,i}\cong P_{d,i}T\mathbb{E}\left[\mathcal{R}_{p,i}\right]+\left(1-P_{d,i}\right)\left(T-\tau_{t,i}\right)\mathbb{E}\left[\mathcal{R}_{p,i}\right].
\end{equation}
Let $\tau_{p,i}=\bar{\mathcal{B}}_{p,i}/\mathbb{E}\left[\mathcal{R}_{p,i}\right]$
where $\tau_{p,i}\in\left[0,\,T\right]$. Higher value of $\tau_{p,i}$ indicates that required average bit-throughput $\bar{\mathcal{B}}_{p,i}$
is higher. Then the QoS constraint $\mathcal{B}_{p,i}\geq\bar{\mathcal{B}}_{p,i}$
can be written as 
\begin{equation}
P_{d,i}\geq\bar{P}_{d,i}=\max\left[0,\,1-\frac{T-\tau_{p,i}}{\tau_{t,i}}\right].\label{eq:Pd_thresh0}
\end{equation}
Thus, to transmit in $i$th channel, detection probability should
be greater than detection probability threshold $\bar{P}_{d,i}$ which
depends on transmission time $\tau_{t,i}$. As $\tau_{t,i}$ increases,
required detection probability increases. To achieve increasing $P_{d,i}$,
sensing time $\tau_{s,i}$ increases, leaving less time available
for transmission. This results in the sensing-throughput tradeoff.
Optimal sensing time is such that constraint in (\ref{eq:Pd_thresh0})
is satisfied with equality \cite{Liang_zheng}.

\subsubsection{Energy constraint}

SU is energy constrained i.e. in each frame, SU has limited energy
to spend in sensing and transmission. This may happen when SU is not
powered by conventional sources and harvests energy from surroundings.
SU employs a greedy policy where it uses all the available energy
in one frame for sensing and transmission subject to maximum power
constraint. Suppose energy $e_{tot}$ is available at SU in each frame.
Then total energy spent in sensing and transmission cannot exceed
$e_{tot}$, that is, 
\begin{equation}
\sum_{i=1}^{M}p_{s}\tau_{s,i}+\sum_{i=1}^{M}p_{t,i}\tau_{t,i}\leq e_{tot}.\label{eq:Energy_constr}
\end{equation}

\subsubsection{Peak transmit power constraint}

As response of power amplifier is non-linear at high values of transmit
power, there a limit $p_{max}$ on allowed maximum transmit power.
Thus, we have 
\begin{equation}
p_{t,i}\leq p_{max},\,\,i=1,\,2,\dots,\,M.\label{eq:Power_constr}
\end{equation}

\subsubsection{Total time constraint}

In a frame, total time spent in sensing and transmission cannot exceed
frame duration. Thus, we have 
\begin{equation}
\sum_{i=1}^{M}\tau_{s,i}+\sum_{i=1}^{M}\tau_{t,i}\leq T.\label{eq:Time_constr}
\end{equation}

\subsection{Problem formulation}

If $i$th channel is vacant, instantaneous rate achieved by SU on
the channel is $\log_{2}\left(1+\frac{g_{i}p_{t,i}}{\sigma_{\mathcal{N}}^{2}}\right)$
bits/s assuming normalized bandwidth. In case of missed detection,
PU interferes with SU transmission. Our interest is in maximizing
throughput achieved in transmission over a vacant band. Thus, we consider
transmission rate achieved under interference as negligible. This
is especially true when interference channel between PU and SU is
strong. Then average bit-throughput of SU over $i$th channel, which
is defined as average number of bits transmitted over $i$th channel
in a frame by SU, is written as 
\begin{equation}
\mathcal{B}_{s,i}=\pi_{0,i}\left(1-P_{f,i}\left(\bar{P}_{d,i},\,\tau_{s,i}\right)\right)\tau_{t,i}\log_{2}\left(1+\frac{g_{i}p_{t,i}}{\sigma_{\mathcal{N}}^{2}}\right).
\end{equation}

In this paper, our objective is to maximize average aggregate bit-throughput
of SU in a frame duration, given by $\mathcal{B}{}_{s}=\sum_{i=1}^{M}\mathcal{B}_{s,i}$
under aforementioned constraints in Section \ref{sub:System-constraints}.
Thus, we can write the maximization problem as

\begin{align}
\max_{\boldsymbol{\tau}_{s},\boldsymbol{\tau}_{t},\boldsymbol{p}_{t}} & \quad\sum_{i=1}^{M}\pi_{0,i}\left(1-P_{f,i}\left(\tau_{t,i},\,\tau_{s,i}\right)\right)\tau_{t,i}\log_{2}\left(1+\frac{g_{i}p_{t,i}}{\sigma_{\mathcal{N}}^{2}}\right)\label{eq:Opt_prob}\\
\mbox{s. t.} & \quad\mbox{(\ref{eq:Pd_thresh0}), (\ref{eq:Energy_constr}), (\ref{eq:Power_constr}), (\ref{eq:Time_constr}),}\nonumber \\
 & \quad\tau_{s,i},\,\tau_{t,i},\,p_{t,i}\geq0,\,\,\,i=1,\,2,\dots,\,M,\nonumber 
\end{align}
where $\boldsymbol{\tau}_{s}=\left[\tau_{s,1},\dots,\,\tau_{s,M}\right]^{T}$,
$\boldsymbol{\tau}_{t}=\left[\tau_{t,1},\dots,\,\tau_{t,M}\right]^{T}$
and $\boldsymbol{p}_{t}=\left[p_{t,1},\,p_{t,2},\dots,\,p_{t,M}\right]^{T}$.
The optimization problem in (\ref{eq:Opt_prob}) is non-convex due
to non-convex nature of $P_{f,i}\left(\tau_{t,i},\,\tau_{s,i}\right)$.
Also, the energy constraint given by (\ref{eq:Energy_constr}) is
non-convex due to product terms of optimization variables $p_{t,i}$
and $\tau_{t,i}$. In following section, we reformulate the problem
so that all constraints are affine and the objective function is separable.

\section{Sensing and resource allocation (SRA): Single user scenario\label{sec:SRA_chap4}}

To make constraint (\ref{eq:Energy_constr}) affine, we reconstitute
problem (\ref{eq:Opt_prob}) as energy and time allocation problem.
Suppose SU uses energy $e_{t,i}$ to transmit in $i$th channel. Then
energy constraint in (\ref{eq:Energy_constr}) can be written as 
\begin{equation}
\sum_{i=1}^{M}e_{t,i}+p_{s}\sum_{i=1}^{M}\tau_{s,i}\leq e_{tot}.\label{eq:Energy_constr_new}
\end{equation}
Transmit power in $i$th channel is $p_{t,i}=e_{t,i}/\tau_{t,i}$.
Thus, we write peak power constraint in (\ref{eq:Power_constr}) as
\begin{equation}
e_{t,i}\leq p_{max}\tau_{t,i},\,\,i=1,\,2,\dots,\,M.\label{eq:Power_constr_new}
\end{equation}

Suppose SU allocates time $t_{s}=\alpha T,\,\alpha\in\left[0,\,1\right]$
for sensing and time $t_{t}=\left(1-\alpha\right)T$ for transmission.
Then we can write time constraint (\ref{eq:Time_constr}) as two separate
constraints given by 
\begin{align}
\sum_{i=1}^{M}\tau_{s,i} & \leq\alpha T,\label{eq:time_constr_s}\\
\sum_{i=1}^{M}\tau_{t,i} & \leq\left(1-\alpha\right)T.\label{eq:time_constr_t}
\end{align}

Let $\mathcal{B}_{s,i}=f_{1,i}\left(\tau_{t,i},\,\tau_{s,i}\right)\cdot f_{2,i}\left(\tau_{t,i},\,e_{t,i}\right)$
where 
\begin{equation}
f_{1,i}\left(\tau_{t,i},\,\tau_{s,i}\right)=1-P_{f,i}\left(\tau_{t,i},\,\tau_{s,i}\right),\label{eq:f_1}
\end{equation}
\begin{equation}
f_{2,i}\left(\tau_{t,i},\,e_{t,i}\right)=\pi_{0,i}\tau_{t,i}\log_{2}\left(1+\frac{g_{i}}{\sigma_{\mathcal{N}}^{2}}\frac{e_{t,i}}{\tau_{t,i}}\right).\label{eq:f_2}
\end{equation}
Now we can reformulate optimization problem in (\ref{eq:Opt_prob})
as follows:

\begin{subequations} 
\begin{align}
\max_{\alpha,\,\left\{ \boldsymbol{\tau}_{s},\,\boldsymbol{\tau}_{t},\,\boldsymbol{e}_{t}\right\} } & \quad\sum_{i=1}^{M}f_{1,i}\left(\tau_{t,i},\,\tau_{s,i}\right)f_{2,i}\left(\tau_{t,i},\,e_{t,i}\right)\label{eq:Opt_prob_new}\\
\mbox{s. t.} & \quad\sum_{i=1}^{M}e_{t,i}+p_{s}\sum_{i=1}^{M}\tau_{s,i}\leq e_{tot},\label{eq:Energy_constr2}\\
 & \quad e_{t,i}\leq p_{max}\tau_{t,i},\,\,i=1,\,2,\dots,\,M,\label{eq:power_constr2}\\
 & \quad\sum_{i=1}^{M}\tau_{s,i}\leq\alpha T,\label{eq:time_constr1}\\
 & \quad\sum_{i=1}^{M}\tau_{t,i}\leq\left(1-\alpha\right)T,\label{eq:time_constr2}\\
 & \quad e_{t,i},\,\tau_{s,i},\,\tau_{t,i}\geq0,\,\,i=1,\,2,\dots,\,M,\label{eq:positive_constr}\\
 & \quad\alpha\in\left[0,\,1\right],
\end{align}
\end{subequations}where $\boldsymbol{\tau}_{s}=\left[\tau_{s,1},\dots,\,\tau_{s,M}\right]^{T}$,
$\boldsymbol{\tau}_{t}=\left[\tau_{t,1},\dots,\,\tau_{t,M}\right]^{T}$
and $\boldsymbol{e}_{t}=\left[e_{t,1},\dots,\,e_{t,M}\right]^{T}$.
In the problem above, all constraints are affine. Objective in (\ref{eq:Opt_prob_new})
is concave in optimization variable $e_{t,i}$. But the problem is
still non-convex in $\tau_{s,i}$ and $\tau_{t,i}$.

To solve (\ref{eq:Opt_prob_new}), we first fix $\alpha$ and decompose
the problem into three subproblems as follows.

\subsection*{Subproblem $\mathbf{P1}$}

We first fix $\left(\boldsymbol{\tau}_{t},\,\boldsymbol{e}_{t}\right)$
and find optimal sensing time $\boldsymbol{\tau}_{s}$ subject to
constraints (\ref{eq:Energy_constr_new}), (\ref{eq:time_constr1})
and (\ref{eq:positive_constr_MU}). Objective in (\ref{eq:Opt_prob_new})
is monotonically increasing with $\tau_{s,i},\,i=1,\,2,\dots,\,M$.
For fixed $\left(\boldsymbol{\tau}_{t},\,\boldsymbol{e}_{t}\right)$,
we can write problem of finding optimal $\boldsymbol{\tau}_{s}$ as

\begin{align}
\max & \quad\sum_{i=1}^{M}\mathbb{\mathcal{B}}_{s,i}\left(\tau_{s,i}\right)\label{eq:subpro1}\\
\mbox{s. t.} & \quad\sum_{i=1}^{M}\tau_{s,i}\leq\min\left[\alpha T,\,\frac{e_{tot}-\sum_{i=1}^{M}e_{t,i}}{p_{s}}\right],\nonumber 
\end{align}
Problem in (\ref{eq:subpro1}) can be modelled as a general non-linear
knapsack problem (NKP). We use greedy algorithm \cite{Hochbaum_knapsack,Zhang_knapsack}
to solve it with complexity $\mathcal{O}\left(M\,\log\left(f_{s}T\right)\right)$.
We define $u_{s}=1/f_{s}$ as the smallest time unit that can be allocated
to the sensing time of a channel. Incidentally $u_{s}$ is also the
time between successive samples. We initialize sensing time as $\tau_{s,i}=0,\,i=1,\,2,\dots,\,M$.
Increase in bit-throughput due to addition of one sensing time unit
can be viewed as reward of the action. Thus, reward of adding a unit
to $i$th channel in $k$th iteration is given by 
\begin{equation}
r_{s,i}^{\left(k\right)}=\mathcal{B}_{s,i}\left(\tau_{s,i}^{\left(k\right)}+u_{s}\right)-\mathcal{B}_{s,i}\left(\tau_{s,i}^{\left(k\right)}\right).
\end{equation}
In each iteration, one time unit is added to the channel $\hat{i}$
where $\hat{i}$ is the channel that gives maximum reward, i.e. $\hat{i}=\arg\,\max_{i}\left\{ r_{s,i}^{\left(k\right)}\right\} $.
Thus, in each iteration, sensing time is updated as 
\begin{equation}
\tau_{s,i}^{\left(k+1\right)}=\begin{cases}
\tau_{s,i}^{\left(k\right)}+u_{s} & \,\,\mbox{for}\,\,i=\hat{i}^{\left(k\right)}\\
\tau_{s,i}^{\left(k\right)} & \,\,\mbox{for}\,\,i\neq\hat{i}^{\left(k\right)}
\end{cases}.
\end{equation}
The process continues until $\sum_{i=1}^{M}\tau_{s,i}^{\left(k\right)}\leq\min\left[\alpha T,\,\frac{e_{tot}-\sum_{i=1}^{M}e_{t,i}}{p_{s}}\right]$.

\subsection*{Subproblem $\mathbf{P2}$}

Keeping optimal $\left(\boldsymbol{\tau}_{s},\,\boldsymbol{e}_{t}\right)$
in $\mathbf{P1}$ fixed, we now optimize $\boldsymbol{\tau}_{t}$
subject to constraints (\ref{eq:time_constr2}) and (\ref{eq:positive_constr}).
The problem of optimizing $\tau_{t,i}$ is 
\begin{align}
\max & \quad\sum_{i=1}^{M}\mathcal{B}_{s,i}\left(\tau_{t,i}\right)\label{eq:subpro2}\\
\mbox{s. t.} & \quad\sum_{i=1}^{M}\tau_{t,i}\leq\left(1-\alpha\right)T.\nonumber 
\end{align}
On similar lines of subproblem $\mathbf{P1}$, we can find optimal
$\tau_{t,i}$ by greedy method for NKP using $u_{t}=1/f_{s}$ as the
smallest time unit. From (\ref{eq:f_1}) and (\ref{eq:f_2}), we see
that $f_{1,i}$ is a monotonically decreasing function of $\tau_{t,i}$
while $f_{2,i}$ is a monotonically increasing function of $\tau_{t,i}$.
Thus, in a region where $\mathcal{B}_{s,i}\left(\tau_{t,i}\right)$
is monotonically decreasing with $\tau_{t,i}$, addition of a unit
$u_{t}$ to transmission time of $i$th channel results in negative
reward value. When reward values for all channel are negative, any
further increase in transmission time $\tau_{t,i}$ results in decreasing
bit-throughput. Thus, the greedy algorithm stops when all rewards
become negative or when constraint (\ref{eq:time_constr2}) is violated.
Using reward $r_{t,i}^{\left(k\right)}=\mathcal{B}_{s,i}\left(\tau_{t,i}^{\left(k\right)}+u_{t}\right)-\mathcal{B}_{s,i}\left(\tau_{t,i}^{\left(k\right)}\right)$,
transmission time is updated in each iteration as 
\[
\tau_{t,i}^{\left(k+1\right)}=\begin{cases}
\tau_{t,i}^{\left(k\right)}+u_{t} & \,\,\mbox{for}\,\,i=\hat{i}^{\left(k\right)}\\
\tau_{t,i}^{\left(k\right)} & \,\,\mbox{for}\,\,i\neq\hat{i}^{\left(k\right)}
\end{cases},
\]
where $\hat{i}^{\left(k\right)}=\arg\,\max_{i}\left\{ r_{t,i}^{\left(k\right)}\right\} $.
The process continues until $\sum_{i=1}^{M}\tau_{t,i}^{\left(k\right)}\leq\left(1-\alpha\right)T$
and $\max\left\{ r_{t,i}^{\left(k\right)}\right\} \geq0$.

\subsection*{Subproblem $\mathbf{P3}$}

Keeping optimal $\left(\boldsymbol{\tau}_{s},\,\boldsymbol{\tau}_{t}\right)$
in $\mathbf{P1}$ and $\mathbf{P2}$ fixed, we now optimize over $\boldsymbol{e}_{t}$
subject to constraints (\ref{eq:Energy_constr2}), (\ref{eq:power_constr2})
and (\ref{eq:positive_constr}). Let $e_{th}=e_{tot}-p_{s}\sum_{i=1}^{M}\left(t_{i}-\tau_{t,i}\right)$.
Since the problem (\ref{eq:Opt_prob_new}) is convex in $e_{t,i}$,
we solve it using Lagrangian method. The Lagrangian for $\mathbf{P3}$
is 
\begin{align}
\mathcal{L}\left(\boldsymbol{e}_{t},\,\lambda,\,\boldsymbol{\mu}\right) & =\sum_{i=1}^{M}f_{1,i}\cdot f_{2,i}\left(e_{t,i}\right)-\lambda\left(\sum_{i=1}^{M}e_{t,i}-e_{th}\right)\nonumber \\
 & \qquad\qquad-\sum_{i=1}^{M}\mu_{i}\left(e_{t,i}-p_{max}\tau_{t,i}\right),
\end{align}
where $\lambda$ and $\boldsymbol{\mu}=\left[\mu_{1},\dots,\mu_{M}\right]^{T}$
denote the dual variables associated with constraints (\ref{eq:Energy_constr2})
and (\ref{eq:power_constr2}). The dual problem of $\mathbf{P3}$
is given by 
\[
\min_{\lambda,\boldsymbol{\mu}}\,\,\max_{\boldsymbol{e}_{t}}\,\,\mathcal{L}.
\]
For fixed $\left(\lambda,\boldsymbol{\mu}\right)$, we find optimal
primal variable by differentiating $\mathcal{L}$ with respect to
$e_{t,i}$ and equating it to zero as 
\begin{equation}
e_{t,i}=\left[\frac{\pi_{0,i}\tau_{t,i}f_{1,i}}{\ln\left(2\right)\left(\lambda+\mu_{i}\right)}-\frac{\sigma_{\mathcal{N}}^{2}\tau_{t,i}}{g_{i}}\right]^{+},
\end{equation}
where $\left[\cdot\right]^{+}=\max\left[\cdot,\,0\right]$. Since
the dual function of $\mathcal{L}$ has unique maximizers, we use
gradient descent method to find $\left(\lambda,\,\boldsymbol{\mu}\right)$
as 
\begin{equation}
\lambda^{\left(k+1\right)}=\lambda^{\left(k\right)}+\epsilon_{\lambda}\left(\sum_{i=1}^{M}e_{t,i}-e_{th}\right),
\end{equation}
\begin{equation}
\mu_{i}^{\left(k+1\right)}=\mu_{i}^{\left(k\right)}+\epsilon_{\mu}\left(e_{t,i}-p_{max}\right),
\end{equation}
where $\epsilon_{\lambda}$ and $\epsilon_{\mu}$ are step sizes.
Iteration index is denoted by $k$. The process of calculating $\boldsymbol{e}_{t}$
and updating $\left(\lambda,\boldsymbol{\mu}\right)$ is repeated
until convergence. In this way the subproblem $\mathbf{P3}$ is solved
with complexity $\mathcal{O}\left(M^{2}\right)$.

All subproblems aim to maximize objective in (\ref{eq:Opt_prob_new}).
For fixed $\alpha$, we repeat the three step process of solving $\mathbf{P1}$,
$\mathbf{P2}$ and $\mathbf{P3}$. This process of finding $\left(\boldsymbol{\tau}_{s},\,\boldsymbol{\tau}_{t},\,\boldsymbol{e}_{t}\right)$
that maximize SU throughput is Block Coordinate Minimization (BCM)
method which converges to stationary solution for non-convex problems
as proven in \cite{BCMconverge_Raza}. Convergence is achieved as
long as initialization of $\left(\boldsymbol{\tau}_{t},\,\boldsymbol{e}_{t}\right)$
in subproblem $\mathbf{P1}$ is done to satisfy constraints (\ref{eq:power_constr2}),
(\ref{eq:time_constr2}) and $\sum_{i=1}^{M}e_{t,i}\leq e_{tot}-p_{s}\alpha T$.
The BCM method runs over all values of $\alpha\in\left[0,\,1\right]$
and value of $\alpha$ that corresponds to the maximum SU bit-throughput
$\mathcal{B}_{s}$ is chosen.

\section{Sensing and resource allocation (SRA): Multi-user scenario\label{sec:SRA_multi_chap4}}

In this section, we propose sensing and resource allocation for the
case where multiple SUs are present in the system. We consider a secondary
network of $N$ SUs governed by a central base station (BS) employing
cooperative sensing. BS acts as the fusion centre for sensing data
of individual SUs. Alternatively, in absence of BS, one of the SUs
can act as the controller. We assume that SUs always have data to
transmit and all SUs transmit to a common destination. To avoid inter-SU
interference, BS employs time division multiple access (TDMA). We
assume that BS has knowledge of channels gains on all SU source to
SU destination links and PU source to SU source links, denoted as
$g_{ij}$ and $h_{ij},\,i\in\left\{ 1,\dots,\,M\right\} ,\,j\in\left\{ 1,\dots,\,N\right\} $
respectively. Assumption of perfect channel knowledge gives us the
upper bound on throughput performance and serves as a baseline for
the case with imperfect or limited channel knowledge. Prior to sensing
and transmission, BS determines optimal sensing time, transmission
time allocation and transmission energy allocation for each channel
and communicates it to the SUs over a low bandwidth control channel
as done in \cite{distributed_Li}.

Time allocated for sensing and transmission in $i$th PU channel is
$\tau_{s,i}$ and $\tau_{t,i}$ respectively. Sensing data is reported
to BS over a low bandwidth control channel. BS performs data fusion
and takes a decision on presence of PU in a given band. In this case,
false-alarm probability in sensing $i$th PU channel is written as \cite{Liang_zheng}
\begin{equation}
P_{f,i}^{'}=\mathcal{Q}\left(\sqrt{2\bar{\gamma}_{i}+1}\mathcal{Q}^{-1}\left(\bar{P}_{d,i}\right)+\bar{\gamma}_{i}\sqrt{f_{s}\tau_{s,i}}\right),\label{eq:coop_false}
\end{equation}
where $\bar{\gamma}_{i}=p_{PU}\sum_{j=1}^{N}h_{ij}$ and $\bar{P}_{d,i}$
is the target detection probability given in (\ref{eq:Pd_thresh0}).

If $i$th PU is sensed to be absent, each SU transmits its own data
to SU destination. Due to TDMA, transmission time of each SU in $i$th
channel is $\tau_{t,i}/N$. Energy used by $j$th SU to transmit in
$i$th channel is $e_{t,ij}$. Energy available at $j$th SU is denoted
by $e_{j}$. A SU participates in the joint-sensing and transmission
process only if it has minimum required energy to sense a channel
for whole frame duration, that is $e_{j}\geq p_{s}T,\,\,j=1,\dots,\,N$.

Our objective is to find optimal access time, transmission time and
energy allocation to maximize sum-throughput of SU system. Let $f_{1,i}^{'}=1-P_{f,i}^{'}$.
Then the optimization problem is

\begin{subequations} 
\begin{align}
\max_{\alpha,\,\left\{ \boldsymbol{\tau}_{s},\,\boldsymbol{\tau}_{t},\,\mathbf{e}_{t}\right\} } & \quad\sum_{i=1}^{M}f_{1,i}^{'}\left(\tau_{t,i},\,\tau_{s,i}\right)\pi_{0,i}\sum_{j=1}^{N}\frac{\tau_{t,i}}{N}\log_{2}\left(1+\frac{g_{ij}N}{\sigma_{\mathcal{N}}^{2}}\frac{e_{t,ij}}{\tau_{t,i}}\right)\label{eq:Opt_prob_MU}\\
\mbox{s. t.} & \quad\sum_{i=1}^{M}e_{t,ij}+p_{s}\sum_{i=1}^{M}\tau_{s,i}\leq e_{j},\,\,j=1,\,2,\dots,\,N,\label{eq:energy_constr_MU}\\
 & \quad e_{t,ij}\leq p_{max}\frac{\tau_{t,i}}{N},\,\,j=1,\,2,\dots,\,N,\label{eq:power_constr_MU}\\
 & \quad\sum_{i=1}^{M}\tau_{s,i}\leq\alpha T,\label{eq:time_constr1_MU}\\
 & \quad\sum_{i=1}^{M}\tau_{t,i}\leq\left(1-\alpha\right)T,\label{eq:time_constr2_MU}\\
 & \quad e_{t,ij},\,t_{i},\,\tau_{t,i}\geq0,\,\,i=1,\dots,\,M,\,j=1,\dots,\,N,\label{eq:positive_constr_MU}\\
 & \quad\alpha\in\left[0,\,1\right],
\end{align}
\end{subequations}where $\boldsymbol{\tau}_{s}=\left[\tau_{s,1},\dots,\,\tau_{s,M}\right]^{T}$,
$\boldsymbol{\tau}_{t}=\left[\tau_{t,1},\dots,\,\tau_{t,M}\right]^{T}$
and $\mathbf{e}_{t}=\left[e_{t,ij}\right]_{M\times N}$. Objective
function (\ref{eq:Opt_prob_MU}) is concave in $\mathbf{e}_{t}$ but
non-convex in $\boldsymbol{\tau}_{s}$ and $\boldsymbol{\tau}_{t}$.
Total energy used by a SU in sensing and transmission cannot exceed
energy available at the SU. This gives rise to a per-user energy constraint
in (\ref{eq:energy_constr_MU}). Constraint in (\ref{eq:power_constr_MU})
is the peak power constraint for each user. Time constraints (\ref{eq:time_constr1_MU})
and (\ref{eq:time_constr2_MU}) remain unchanged from single-user
scenario. We see that all constraints are affine.

Along the lines of Section \ref{sec:SRA_chap4}, we can decompose
the optimization problem in three subproblems for fixed value of $\alpha$.
Subproblem $\mathbf{P1}$ solves sensing time allocation for fixed
$\left(\boldsymbol{\tau}_{t},\,\mathbf{e}_{t}\right)$ using greedy
algorithm for NKP under constraint $\sum_{i=1}^{M}\tau_{s,i}\leq\min\left[\alpha T,\,\tau_{th,1},\,\tau_{th,2},\,\dots,\,\tau_{th,N}\right]$,
where $\tau_{th,j}$ is given by 
\[
\tau_{th,j}=\frac{e_{j}-\sum_{i=1}^{M}e_{t,ij}}{p_{s}},\,\,j=1,\,2,\dots,\,N.
\]
In subproblem $\mathbf{P2}$, to find optimal $\boldsymbol{\tau}_{t}$,
let $f_{2,i}^{'}=\pi_{0,i}\sum_{j=1}^{N}\frac{\tau_{t,i}}{N}\log_{2}\left(1+\frac{g_{ij}N}{\sigma_{\mathcal{N}}^{2}}\frac{e_{t,ij}}{\tau_{t,i}}\right).$
We see that $f_{1,i}^{'}$ is monotonically decreasing function of
$\tau_{t,i}$ and $f_{2,i}^{'}$ is a monotonically increasing function
of $\tau_{t,i}$. Thus, for fixed value of $\left(\boldsymbol{\tau}_{s},\,\mathbf{e}_{t}\right)$,
optimal $\boldsymbol{\tau}_{t}$ can be found by greedy algorithm
with modified stopping criteria as done in subproblem $\mathbf{P2}$
in Section \ref{sec:SRA_chap4}.

In subproblem $\mathbf{P3}$, we keep optimal $\left(\boldsymbol{\tau}_{s},\,\boldsymbol{\tau}_{t}\right)$
fixed and optimize over $\mathbf{e}_{t}$ under constraints (\ref{eq:energy_constr_MU}),
(\ref{eq:power_constr_MU}) and (\ref{eq:positive_constr_MU}). $\mathbf{P3}$
is a convex-programming problem that is solved by Lagrangian method
using steps similar to those used in Section \ref{sec:SRA_chap4}.
We omit the steps here for brevity and write closed form expression
for $e_{t,ij}$ as 
\begin{equation}
e_{t,ij}=\left[\frac{\pi_{0,i}\tau_{t,i}f_{1,i}^{'}}{\ln\left(2\right)\left(\lambda_{j}+\mu_{ij}\right)N}-\frac{\sigma_{\mathcal{N}}^{2}\tau_{t,i}}{g_{ij}N}\right]^{+},
\end{equation}
where $\lambda_{j}$ and $\mu_{ij}$ are Lagrange's multipliers that
are chosen to satisfy per-user energy constraint $\sum_{i=1}^{M}e_{t,ij}\leq e_{j}-p_{s}\sum_{i=1}^{M}\tau_{s,i}$
and peak power constraint $e_{t,ij}\leq p_{max}\frac{\tau_{t,i}}{N}$
respectively. Subproblems $\mathbf{P1}$, $\mathbf{P2}$ and $\mathbf{P3}$
are executed recursively until all variables converge. The process
of finding optimal $\left(\boldsymbol{\tau}_{s},\,\boldsymbol{\tau}_{t},\,\mathbf{e}_{t}\right)$
is repeated over all values of $\alpha\in\left[0,\,1\right]$ and
$\alpha$ that maximizes SU bit-throughput is chosen.

\section{Simulation results and discussion\label{sec:Simulation_chap4}}

In this section, we first study performance of proposed Sensing and
Resource Allocation (SRA) method under different channel conditions
and energy availability scenarios for single-SU case. We compare the
performance with Best Channel Selection (BCS) and Proportional Energy
and Time Allocation (PETA) methods. In BCS, SU chooses the best channel
for sensing and transmission, based on a heuristic that depends on
channel gains, PU occupancy and QoS constraint. The heuristic $\mathcal{H}_{i}$
for $i$th channel is defined as 
\begin{equation}
\mathcal{H}_{i}=\frac{\pi_{0,i}g_{i}}{\tau_{p,i}}.
\end{equation}
Value of $\mathcal{H}_{i}$ is high for a channel with low occupancy
probability, low value of $\tau_{p}$ and good SU-SU channel. In each
frame, SU chooses the best channel $\hat{i}$ that has highest value
of $\mathcal{H}_{i}$, i.e. $\hat{i}=\arg\,\max_{i\in\left\{ 1,\dots,\,M\right\} }\,\mathcal{H}_{i}$.
In PETA, total available time and energy is divided between channels
such that time and energy for each channel is proportional to the
channel heuristic. Let $T_{i}$ and $e_{tot,i}$ be the time allocated
to $i$th channel. Then we have $T_{i}=k\mathcal{H}_{i}T$ and $e_{tot,i}=k\mathcal{H}_{i}e_{tot}$
where normalization factor $k$ is calculated as $k=1/\sum_{i=1}^{M}\mathcal{H}_{i}$.
For each channel, optimal sensing and transmission time $\tau_{s,i},\,\tau_{t,i}\leq T_{i}$
as well as transmission energy $e_{t,i}\leq e_{tot,i}$ is found using
SRA. We also compare the performance with single channel transmission
scheme (SRA-SC). In SRA-SC, expected throughput is calculated for
a single channel at a time. Only the channel that gives maximum expected
throughput is selected for transmission. Further, as a baseline for
performance comparison, we use optimal sensing under target detection
probability $P_{d}=0.95$ proposed in \cite{Liang_zheng} combined
with best channel selection.

For simulation, we consider $M=10$. Values of frame time and sampling
frequency are $T=100\,\mbox{ms}$ and $f_{s}=1\,\mbox{MHz}$. As all
channels are Rayleigh faded, SU-SU channel gains $g_{i},\,\,i=1,\dots,\,M$
are exponentially distributed. We take the average channel gain $\sigma_{g}^{2}=-10\,\mbox{dB}$
unless mentioned otherwise. Channel occupancy probabilities are $\boldsymbol{\pi}_{1}=\left[0.7,0.7,0.7,0.7,0.7,0.5,0.5,0.5,0.5,0.5\right]^{T}$.
PU QoS threshold for each PU channel is $\tau_{p,i}=0.9,\,i=1,\dots,\,M$.
Maximum power threshold is $p_{max}=1\,\mbox{W}$ and power required
for sensing is $p_{s}=0.1\,\mbox{W}$ \cite{EH_Park}. Noise power
is $\sigma_{\mathcal{N}}^{2}=0.1\,\mbox{W}$. Average received PU
SNR is $\gamma_{i}=\gamma_{p}=-10\,\mbox{dB},\,\,i=1,\dots,\,M$,
unless mentioned otherwise.

\begin{figure}
\centering

\includegraphics[scale=0.6]{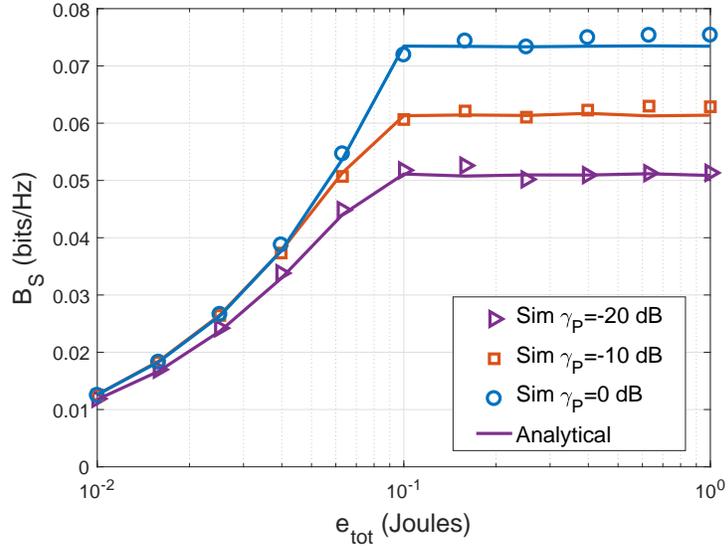}

\protect\protect\protect\caption{Bit-throughput $\mathcal{B}_{s}$ versus available energy $e_{tot}$
for different values of average received PU SNR $\gamma_{p}$ for
$N=1$. \label{fig:CapVsE_SNR}}
\end{figure}

\begin{figure}
\centering

\includegraphics[scale=0.6]{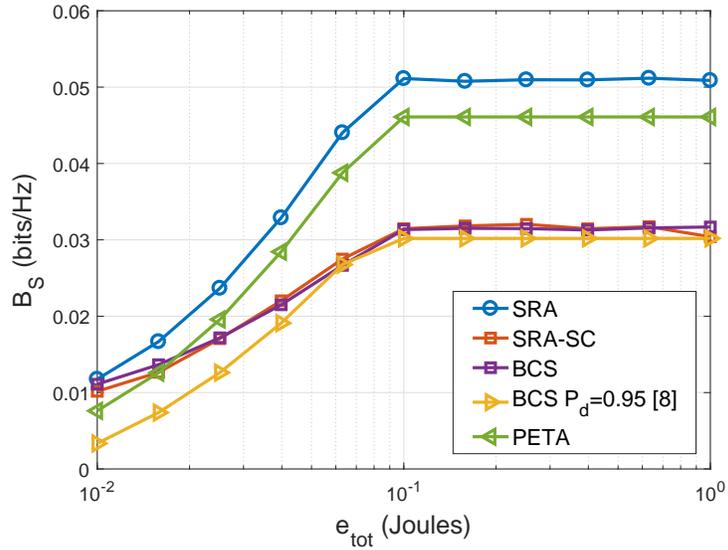}

\protect\protect\protect\caption{Comparison of different allocation methods in terms of bit-throughput
$\mathcal{B}_{s}$ for $N=1$ and $\gamma_{P}=-10\,\mbox{dB}$.\label{fig:CapVsE_Compare}}
\end{figure}

\subsection{Effect of energy availability}

Fig. \ref{fig:CapVsE_SNR} plots simulation and analytical results
for bit-throughput achieved in proposed SRA method against available
energy for different values of received PU power. As available energy
$e_{tot}$ increases, more energy can be used in transmission and
average bit-throughput $\mathcal{B}_{s}$ increases. At high value
of $e_{tot}$, peak power constraint in (\ref{eq:power_constr2})
becomes dominant and limits transmission energy in each channel. Even
though energy is available, more energy cannot be used in transmitting.
Thus, $\mathcal{B}_{s}$ becomes constant at high value of $e_{tot}$.

When received PU power is high, sensing time required to achieve target
detection probability is low. Also, false alarm probability is low.
This leaves more time for transmission in a channel. Thus, throughput
achieved is higher. If received PU power is low, more sensing time
is required to detect a PU correctly. Thus, time available for transmission
decreases, resulting in less bit-throughput.

Fig. \ref{fig:CapVsE_Compare} shows that bit-throughput in SRA is
better than BCS, PETA and SRA-SC. In BCS, best channel chosen by SU
may have desirable properties like low occupancy and loose QoS constraint
but may also have low channel gain. By choosing a single channel to
transmit, diversity provided by multiple channels is sacrificed in
BCS, hence resulting in less throughput. PETA fixes maximum energy
and time allocated to a channel based on channel heuristics and only
optimizes over individual channels, which is suboptimal compared to
SRA. Fig. \ref{fig:CapVsE_Compare} also plots throughput for optimal
sensing in \cite{Liang_zheng} using best channel selection. We see
that SRA performs better than fixed $P_{d}$ based method.

\subsection{Effect of primary occupancy and QoS constraint}

\begin{figure}
\centering

\includegraphics[scale=0.6]{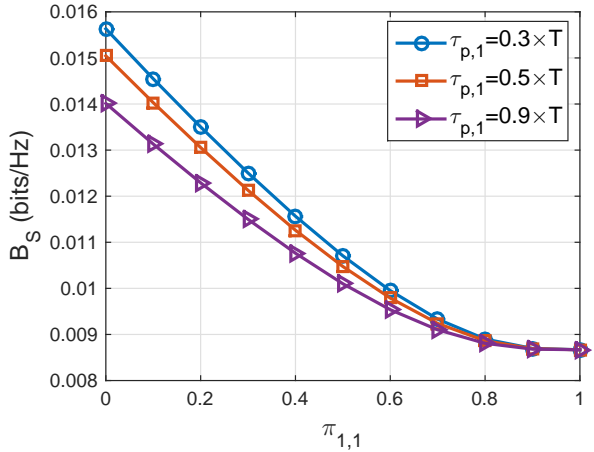}

\protect\protect\protect\caption{Bit-throughput $\mathcal{B}_{s}$ versus PU occupancy probability
$\pi_{1,1}$ for different values of QoS constraint $\tau_{p,1}$\label{fig:CapVsPi}
for $e_{tot}=10\,\mbox{mJ}$, $M=4$ and $N=1$.}
\end{figure}

Fig. \ref{fig:CapVsPi} plots bit-throughput $\mathcal{B}_{s}$ against
occupancy probability $\pi_{1,1}$ occupancy probabilities of other
channels unchanged. If channel gain of the SU-SU link is high, with
increasing occupancy probability $\pi_{1,i}$ more sensing time has
to be allotted to the channel before transmission. This results in
reduced overall average throughput. Similarly, as QoS constraint $\tau_{p,1}$
becomes more stringent, target detection probability increases, requiring
more sensing time. This leaves less time for transmission and results
in reduced throughput. Thus, with increasing PU occupancy probability
and tighter QoS constraint, bit-throughput of SU decreases.

\begin{figure}
\centering

\includegraphics[scale=0.5]{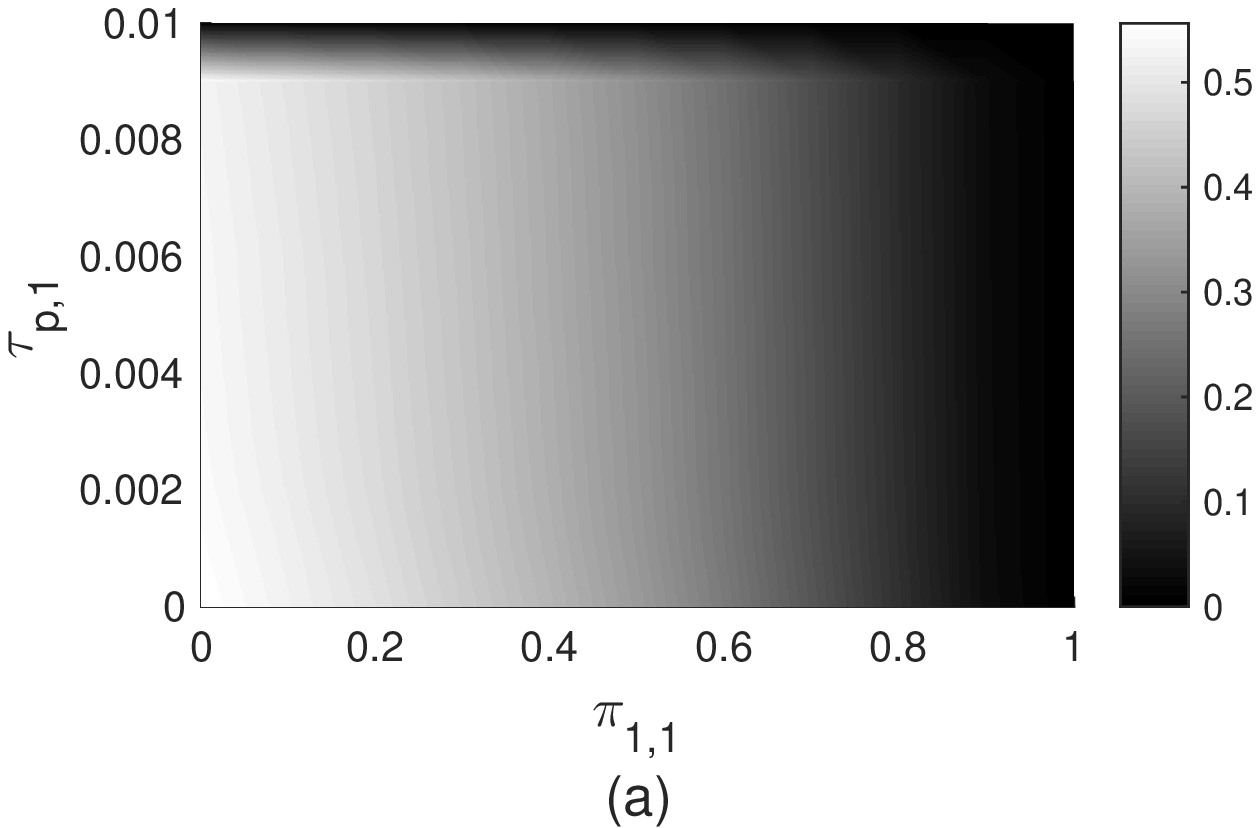}$\!\!\!\!\!$\includegraphics[scale=0.5]{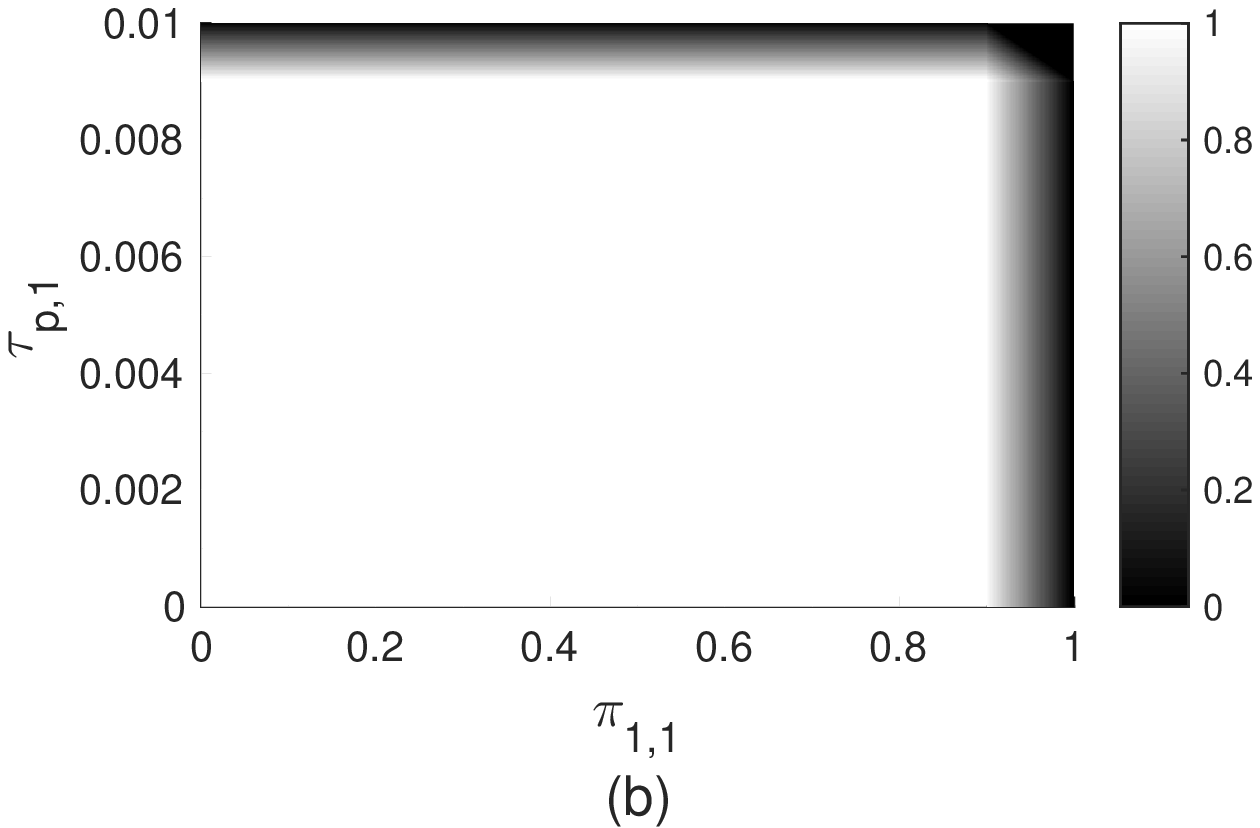}

\protect\protect\protect\caption{Channel access probability versus PU occupancy probability $\pi_{1,1}$
and QoS constraint $\tau_{p,1}$ for $T=10\,\mbox{ms}$, $f_{s}=0.1\,\mbox{MHz}$,
$M=4$ and (a) $e_{tot}=10^{-4}\,\mbox{J}$ (b) $e_{tot}=10^{-2}\,\mbox{J}$.\label{fig:ProbVsPi}}
\end{figure}

PU occupancy and QoS constraints also affect probability of SU accessing
a channel as shown in Fig. \ref{fig:ProbVsPi}. Channel access probability
is defined as probability of SU transmitting in the channel. It is
represented in Fig. \ref{fig:ProbVsPi} in grayscale color tone where
darker shade indicates lower access probability. As $\pi_{1,1}$ and
$\tau_{p,1}$ increase, SU does not transmit in the channel unless
SU-SU channel gain is high. Thus, channel access probability decreases.
When available energy $e_{tot}$ is high, maximum power constraint
limits the energy that can be transmitted in a channel. Thus, some
energy is distributed in other channels even though their occupancy
probability may be higher and QoS constraints may be tighter. Thus,
overall access probability is higher than low energy availability
case. Access probability decreases sharply only at high values of
$\pi_{1,i}$ and $\tau_{p,i}$ as seen in Fig. \ref{fig:ProbVsPi}(b).

\subsection{Sum-throughput in multi-user scenario}

\begin{figure}
\centering

\includegraphics[scale=0.6]{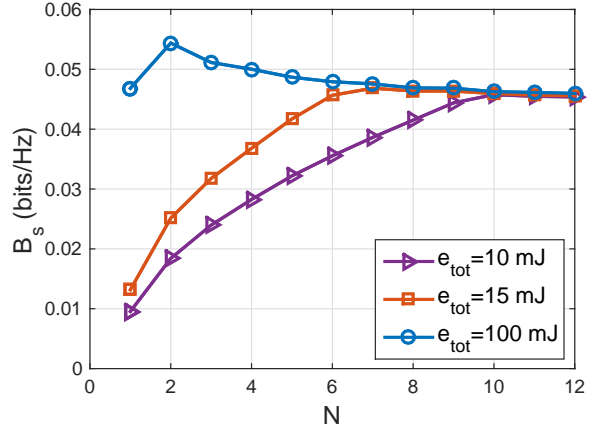}

\protect\protect\protect\caption{Sum-throughput $\mathcal{B}_{s}$ versus number of SUs $N$ for different
values of available energy $e_{tot}$ for $M=4$.\label{fig:CapVsN}}
\end{figure}

Heuristic based methods are suboptimal and SRA clearly outperforms
them as seen in Fig. \ref{fig:CapVsE_Compare}. Also, for multi-user
scenario, allocating resources for each SU-PU channel pair based on
heuristics is a separate optimization problem in itself. Hence, we
omit the comparison of SRA with heuristic based methods. In Fig. \ref{fig:CapVsN},
we plot sum-throughput of a SU network against number of SUs $N$
for different energy availability scenarios. Energy available at each
SU is denoted by $e_{tot}$. Initially, as number of SUs increase,
cooperative sensing lowers required sensing time to achieve target
detection probability leaving more time for transmission. This results
in increasing sum-throughput. But increasing number of SUs decrease
transmission time allotted to each SU, given by $\frac{\tau_{t,i}}{N},\,i=1,\dots,\,M$.
Due to peak power constraint $e_{t,ij}\leq p_{max}\frac{\tau_{t,i}}{N}$,
transmission energy decreases with decreasing value of $\frac{\tau_{t,i}}{N}$.
Thus, with increasing $N$, transmission time as well as transmission
energy of each SU decreases, resulting in decreasing throughput.

\section{Conclusion\label{sec:Conclusion_chap4}}

We considered a CR system with multiple PU channels where simultaneous
sensing of all channels is not possible. Also, SU has limited energy
for sensing and transmission. The problem of maximizing SU bit-throughput
while maintaining QoS of PUs was formulated as a resource allocation
problem with time and energy as available resources. We proposed sensing
and resource allocation (SRA) method that solves the problem by decomposing
it in non-linear knapsack subproblem and convex optimization subproblem.
We then extended the framework for multiple SU scenario where SUs
can achieve benefits of cooperative sensing. Simulation results show
that throughput increases as more energy is available. It was observed
that with more number of SUs, throughput performance benefits from
cooperative sensing. But as SUs further increase, throughput decreases
as less time is available for transmission of each SU.


\begingroup \let\itshape\upshape  
\bibliographystyle{ieeetr}
\bibliography{database_full}

\begin{thebibliography}{10}

\bibitem{FCC}
M.~A. McHenry, ``{NSF spectrum occupancy measurements project summary},'' {\em
  Shared Spectrum Company}, 2005.

\bibitem{Haykin}
S.~Haykin, ``Cognitive radio: brain-empowered wireless communications,'' {\em
  IEEE J. Sel. Areas Commun.}, vol.~23, pp.~201--220, Feb 2005.

\bibitem{Yucek}
T.~Yucek and H.~Arslan, ``A survey of spectrum sensing algorithms for cognitive
  radio applications,'' {\em IEEE Commun. Surveys Tuts.}, vol.~11,
  pp.~116--130, First 2009.

\bibitem{Sens_survey_Sun}
H.~Sun, A.~Nallanathan, C.-X. Wang, and Y.~Chen, ``Wideband spectrum sensing
  for cognitive radio networks: a survey,'' {\em IEEE Wireless Commun.},
  vol.~20, pp.~74--81, April 2013.

\bibitem{Digham_energy_det}
F.~Digham, M.-S. Alouini, and M.~K. Simon, ``On the energy detection of unknown
  signals over fading channels,'' {\em IEEE Trans. Commun.}, vol.~55,
  pp.~21--24, Jan 2007.

\bibitem{Urkowitz_energy_det}
H.~Urkowitz, ``Energy detection of unknown deterministic signals,'' {\em Proc.
  IEEE}, vol.~55, pp.~523--531, April 1967.

\bibitem{Goldsmith2}
A.~Goldsmith, S.~Jafar, I.~Maric, and S.~Srinivasa, ``Breaking spectrum
  gridlock with cognitive radios: An information theoretic perspective,'' {\em
  Proc. IEEE}, vol.~97, pp.~894--914, May 2009.

\bibitem{Liang_zheng}
Y.-C. Liang, Y.~Zeng, E.~Peh, and A.~T. Hoang, ``Sensing-throughput tradeoff
  for cognitive radio networks,'' {\em IEEE Trans. Wireless Commun.}, vol.~7,
  pp.~1326 --1337, april 2008.

\bibitem{sens_tradeoff_Peh}
E.~Peh, Y.-C. Liang, Y.~L. Guan, and Y.~Zeng, ``Optimization of cooperative
  sensing in cognitive radio networks: A sensing-throughput tradeoff view,''
  {\em IEEE Trans. Veh. Technol.}, vol.~58, pp.~5294--5299, Nov 2009.

\bibitem{sens_tradeoff_zarrin}
S.~Zarrin and T.~J. Lim, ``Throughput-sensing tradeoff of cognitive radio
  networks based on quickest sensing,'' in {\em IEEE Int. Conf. Commun. (ICC
  2011)}, pp.~1--5, June 2011.

\bibitem{Sens_thru_Juarez}
M.~Cardenas-Juarez and M.~Ghogho, ``Spectrum sensing and throughput trade-off
  in cognitive radio under outage constraints over {Nakagami} fading,'' {\em
  IEEE Commun. Lett.}, vol.~15, pp.~1110--1113, October 2011.

\bibitem{Sens_tradeoff_Kaushik}
A.~Kaushik, S.~Sharma, S.~Chatzinotas, B.~Ottersten, and F.~Jondral,
  ``Sensing-throughput tradeoff for cognitive radio systems with unknown
  received power,'' in {\em Cognitive Radio Oriented Wireless Networks}
  (M.~Weichold, M.~Hamdi, M.~Z. Shakir, M.~Abdallah, G.~K. Karagiannidis, and
  M.~Ismail, eds.), vol.~156 of {\em Lecture Notes of the Institute for
  Computer Sciences, Social Informatics and Telecommunications Engineering},
  pp.~308--320, Springer International Publishing, 2015.

\bibitem{sens_thru_Pei}
Y.~Pei, Y.-C. Liang, K.~Teh, and K.~H. Li, ``Sensing-throughput tradeoff for
  cognitive radio networks: A multiple-channel scenario,'' in {\em Proc. IEEE
  20th Int. Symp. Personal, Indoor and Mobile Radio Commun. (PIMRC 2009)},
  pp.~1257--1261, Sept 2009.

\bibitem{sens_tradeoff_sharkasi}
Y.~Sharkasi, M.~Ghogho, and D.~McLernon, ``Sensing-throughput tradeoff for
  {OFDM}-based cognitive radio under outage constraints,'' in {\em Proc. Int.
  Symp. Wireless Commun. Syst. (ISWCS 2012)}, pp.~66--70, Aug 2012.

\bibitem{Paradiso}
J.~Paradiso and T.~Starner, ``Energy scavenging for mobile and wireless
  electronics,'' {\em IEEE Pervasive Comput.}, vol.~4, pp.~18--27, Jan 2005.

\bibitem{EH_Lu}
X.~Lu, P.~Wang, D.~Niyato, D.~I. Kim, and Z.~Han, ``Wireless networks with {RF}
  energy harvesting: {A} contemporary survey,'' {\em CoRR}, vol.~abs/1406.6470,
  2014.

\bibitem{EH_Park}
S.~Park, H.~Kim, and D.~Hong, ``Cognitive radio networks with energy
  harvesting,'' {\em IEEE Trans. Wireless Commun.}, vol.~12, pp.~1386--1397,
  March 2013.

\bibitem{sens_eff_Wang}
S.~Wang, Y.~Wang, J.~Coon, and A.~Doufexi, ``Energy-efficient spectrum sensing
  and access for cognitive radio networks,'' {\em IEEE Trans. Veh. Technol.},
  vol.~61, pp.~906--912, Feb 2012.

\bibitem{sens_E_tradeoff_Yin}
S.~Yin, E.~Zhang, L.~Yin, and S.~Li, ``Saving-sensing-throughput tradeoff in
  cognitive radio systems with wireless energy harvesting,'' in {\em Proc. IEEE
  Global Commun. Conf. (GLOBECOM 2013)}, pp.~1032--1037, Dec 2013.

\bibitem{sens_eff_Wu}
X.~Wu, J.-L. Xu, M.~Chen, and J.~Wang, ``Optimal energy-efficient sensing and
  power allocation in cognitive radio networks,'' 2014.

\bibitem{sens_eff_Hoang}
A.~T. Hoang, Y.-C. Liang, D.~Tung Chong~Wong, R.~Zhang, and Y.~Zeng,
  ``Opportunistic spectrum access for energy-constrained cognitive radios,'' in
  {\em Proc. IEEE Veh. Technol. Conf. (VTC Spring 2008)}, pp.~1559--1563, May
  2008.

\bibitem{Sens_Eff_Sultan}
A.~Sultan, ``Sensing and transmit energy optimization for an energy harvesting
  cognitive radio,'' {\em IEEE Wireless Commun. Lett.}, vol.~1, pp.~500--503,
  October 2012.

\bibitem{sens_eff_Park}
S.~Park, S.~Lee, B.~Kim, D.~Hong, and J.~Lee, ``Energy-efficient opportunistic
  spectrum access in cognitive radio networks with energy harvesting,'' in {\em
  Proceedings of the 4th International Conference on Cognitive Radio and
  Advanced Spectrum Management}, CogART '11, (New York, NY, USA),
  pp.~62:1--62:5, ACM, 2011.

\bibitem{Spectrum_survey_Chakraborty}
A.~Chakraborty and S.~R. Das, ``Measurement-augmented spectrum databases for
  white space spectrum,'' in {\em Proceedings of the 10th ACM International on
  Conference on Emerging Networking Experiments and Technologies}, CoNEXT '14,
  (New York, NY, USA), pp.~67--74, ACM, 2014.

\bibitem{Spectrum_survey_Das}
D.~Das and S.~Das, ``A survey on spectrum occupancy measurement for cognitive
  radio,'' {\em Wireless Personal Communications}, vol.~85, no.~4,
  pp.~2581--2598, 2015.

\bibitem{Spectrum_survey_Lehtomaki}
J.~J. Lehtomaki, R.~Vuohtoniemi, and K.~Umebayashi, ``On the measurement of
  duty cycle and channel occupancy rate,'' {\em IEEE J. Sel. Areas Commun.},
  vol.~31, pp.~2555--2565, November 2013.

\bibitem{Spectrum_survey_Xue}
J.~Xue, Z.~Feng, and P.~Zhang, ``Spectrum occupancy measurements and analysis
  in {Beijing},'' {\em \{IERI\} Procedia}, vol.~4, pp.~295 -- 302, 2013.
\newblock 2013 International Conference on Electronic Engineering and Computer
  Science (EECS 2013).

\bibitem{Zhao}
Q.~Zhao, L.~Tong, A.~Swami, and Y.~Chen, ``Decentralized cognitive {MAC} for
  opportunistic spectrum access in ad hoc networks: A {POMDP} framework,'' {\em
  IEEE J. Sel. Areas Commun.}, vol.~25, pp.~589--600, April 2007.

\bibitem{Spectrum_karandikar}
G.~Naik, S.~Singhal, A.~Kumar, and A.~Karandikar, ``Quantitative assessment of
  {TV} white space in india,'' in {\em Proc. National Conf. Commun. (NCC
  2014)}, pp.~1--6, Feb 2014.

\bibitem{Hochbaum_knapsack}
D.~S. Hochbaum, ``A nonlinear knapsack problem,'' {\em Operations Research
  Letters}, pp.~103--110, 1995.

\bibitem{Zhang_knapsack}
Y.~Zhang and C.~Leung, ``Subcarrier, bit and power allocation for multiuser
  {OFDM}-based multi-cell cognitive radio systems,'' in {\em Proc. IEEE Veh.
  Technol. Conf. (VTC 2008-Fall)}, pp.~1--5, Sept 2008.

\bibitem{BCMconverge_Raza}
M.~Razaviyayn, M.~Hong, and Z.-Q. Luo, ``A unified convergence analysis of
  block successive minimization methods for nonsmooth optimization,'' {\em SIAM
  Journal on Optimization}, vol.~23, no.~2, pp.~1126--1153, 2013.

\bibitem{distributed_Li}
T.~Li, N.~B. Mandayam, and A.~Reznik, ``A framework for distributed resource
  allocation and admission control in a cognitive digital home,'' {\em IEEE
  Trans. Wireless Commun.}, vol.~12, pp.~984--995, March 2013.

\end{thebibliography}
 \endgroup



\parpic{\includegraphics[scale=0.8,clip,keepaspectratio]{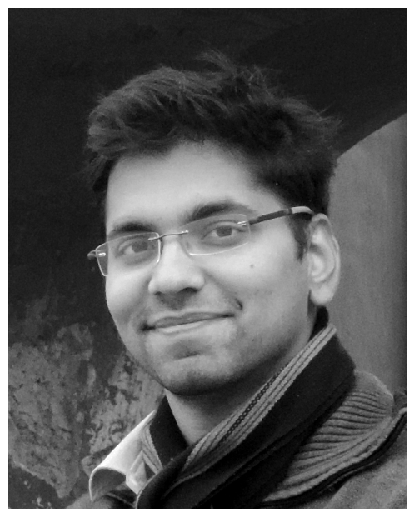}}
\noindent {\bf Kedar Kulkarni} received his B.Tech. degree in Electronics
and Telecommunication from College of Engineering Pune,
India, in 2010. He is presently pursuing Ph.D. degree in
Department of Electrical Engineering at Indian Institute of
Technology (IIT) Kanpur, India. His research interests
include resource allocation in cognitive radio networks,
cooperative networks, and energy harvesting wireless
communications. \\ \\

\parpic{\includegraphics[scale=0.8,clip,keepaspectratio]{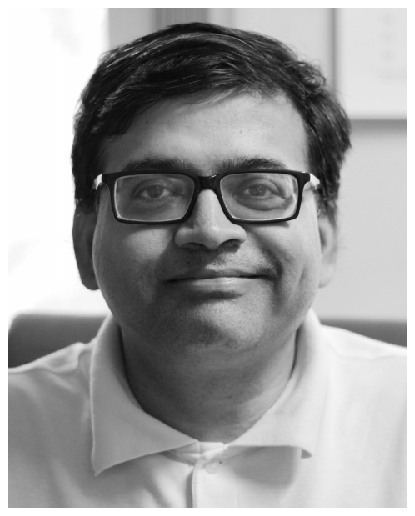}}
\noindent {\bf Adrish Banerjee} received his B.Tech. degree in
Electronics and Electrical Communication Engineering
from Indian Institute of Technology (IIT) Kharagpur, India,
in 1996 and M.S. and Ph.D. degrees in Electrical
Engineering from University of Notre Dame, Indiana in
1998 and 2003, respectively. He is currently an Associate
Professor in the Department of Electrical Engineering at IIT
Kanpur, India. His research interests include physical layer
aspects of wireless communications, particularly error
control coding, cognitive radio, and green communications.

\end{document}